\documentclass[a4paper]{PoS}
\usepackage{graphicx}
\usepackage{caption}

\title{Silicon Photomultiplier Research and Development Studies for the Large Size Telescope of the Cherenkov Telescope Array}

\ShortTitle{SiPM R\&D Studies for the Large Size Telescope of CTA}

\author{\speaker{Riccardo Rando}$^{ab}$, Daniele Corti$^a$, Francesco Dazzi$^c$, Alessandro De Angelis$^d$, Antonios Dettlaff$^c$, Daniela Dorner$^e$, David Fink$^c$, Nadia Fouque$^f$, Felix Grundner$^c$, Werner Haberer$^c$, Alexander Hahn$^c$, Richard Hermel$^f$, Samo Korpar$^g$, Ga\v{s}per Kukec Mezek$^h$, Ronald Maier$^c$, Christian Manea$^a$, Mos\`e Mariotti$^{ab}$, Daniel Mazin$^{ci}$, Fatima Mehrez$^f$, Razmik Mirzoyan$^c$, Sergey Podkladkin$^c$, Ignasi Reichardt$^a$, Wolfgang Rhode$^j$, Sylvie Rosier$^f$, Cornelia Schultz$^{ab}$, Carlo Stella$^k$, Masahiro Teshima$^{ci}$, Holger Wetteskind$^c$ and Marko Zavrtanik$^{gh}$ for the CTA Consortium\footnote{Full Consortium author list at http://cta-observatory.org}\\
E-mail: \email{riccardo.rando@pd.infn.it}

{\footnotesize 
{$^a$}~INFN Section of Padova, Italy; 
{$^b$}~University of Padova, Italy; 
{$^c$}~Max Planck Institute for Physics, Munich, Germany; 
{$^d$}~LIP/IST Lisboa, Portugal; 
{$^e$}~University of W\"{u}rzburg, Germany; 
{$^f$}~LAPP, CNRS-IN2P3, Annecy, France; 
{$^g$}~Jo\v{z}ef Stefan Institute, Ljubljana, Slovenia; 
{$^h$}~University of Nova Gorica, Slovenia; 
{$^i$}~University of Tokyo and ICRR, Tokyo, Japan; 
{$^j$}~Technische Universit\"at, Dortmund, Germany; 
{$^k$}~University of Udine and INFN, Italy.
}

}

\abstract{The Cherenkov Telescope Array (CTA) is the the next generation facility of imaging atmospheric Cherenkov telescopes; two sites will cover both hemispheres. CTA will reach unprecedented sensitivity, energy and angular resolution in very-high-energy gamma-ray astronomy. Each CTA array will include four Large Size Telescopes (LSTs), designed to cover the low-energy range of the CTA sensitivity ($\sim$20~GeV to 200~GeV). In the baseline LST design, the focal-plane camera will be instrumented with 265 photodetector clusters; each will include seven photomultiplier tubes (PMTs), with an entrance window of 1.5~inches in diameter. The PMT design is based on mature and reliable technology. Recently, silicon photomultipliers (SiPMs) are emerging as a competitor. Currently, SiPMs have advantages (e.g. lower operating voltage and tolerance to high illumination levels) and disadvantages (e.g. higher capacitance and cross talk rates), but this technology is still young and rapidly evolving. SiPM technology has a strong potential to become superior to the PMT one in terms of photon detection efficiency and price per square mm of detector area. While the advantage of SiPMs has been proven for high-density, small size cameras, it is yet to be demonstrated for large area cameras such as the one of the LST. We are working to develop a SiPM-based module for the LST camera, in view of a possible camera upgrade. We will describe the solutions we are exploring in order to balance a competitive performance with a minimal impact on the overall LST camera design.}

\FullConference{The 34th International Cosmic Ray Conference,\\
		30 July- 6 August, 2015\\
		The Hague, The Netherlands}

\begin{document}

\section{The Cherenkov Telescope Array and the Large Size Telescope}

A ground-based IACT measures the Cherenkov light from an air shower generated by the interaction of very energetic particles from the Universe in the upper atmosphere, looking in particular at cosmic accelerators emitting high-energy gamma rays.

Following the great success of the current-generation IACTs, the preparation of the next generation very-high-energy gamma-ray observatory CTA is under way \cite{ref:cta}. Two observatories are planned, one in the Northern hemisphere and the other in the Southern one. Four Large Size Telescopes (LST), 23~m in diameter and with a focal length of 28~m, are arranged at the centre of both arrays, to lower the energy threshold and to improve the sensitivity of CTA between 20 and 200~GeV \cite{ref:lstmain}. A reflective surface of 368~m$^2$ collects and focuses the Cherenkov radiation into the camera, where 1855 photosensors convert the light into electrical signals that can be processed by dedicated electronics. The camera has been designed for maximum compactness and lowest weight, cost and power consumption. In the current design, each pixel incorporates a photomultiplier tube (PMT) and the corresponding readout electronics \cite{ref:LSTelec}. The selected PMTs have an input window of 1.5~inch in diameter and are optimized for a high photodetection efficiency in the near-UV region, where the spectrum of the Cherenkov light peaks. In front of each PMT, a reflective-foil light concentrator compresses the light coming from the main mirror by a factor $\sim4$ in area.

\section{SiPM R\&D for the LST camera}

Silicon photomultipliers (SiPMs) are very promising sensors for Cherenkov applications, thanks to the high quantum efficiency, high gain, fast response, and low-amplitude afterpulses. On the other hand, the relatively high dark count rate is mostly irrelevant, due to the large background photon flux from the night sky. In fact, the feasibility of SiPM for IACT applications has been demonstrated for small scale telescopes \cite{ref:fact1,ref:fact2}; in addition, the First G-APD Cherenkov Telescope (FACT) has shown that SiPMs have a stable and reliable performance on the long term \cite{ref:fact3}. Within CTA, the smaller telescopes will use this technology \cite{ref:sct-sipm,ref:sct-sipm2,ref:sct-sipm3,ref:infn-camera}.

In the case of LST the pixel size is significantly larger, and this adds additional complications. Nonetheless, the advantages of SiPM sensors, and the rapid pace at which this technology is progressing, warrant an effort to investigate a possible SiPM-based design for the LST camera. The CTA LST SiPM Research \& Development working group has the goal of providing a demonstrator unit for a possible LST camera upgrade, with scientific performance equal or superior to the PMT-based design. To ensure the economical feasibility of the project, one additional requirement is to keep most of the existing camera readout \cite{ref:lst-pmt-ro}, so a complete optimization is not considered. The approach described here involves covering the pixel's active area with several SiPM sensors, and summing the analog signals into one single output, to be processed by the existing PMT readout chain. Single-photon spectroscopy, i.e. the capability of clearly separating the 1-photon peak is not strictly necessary for the scientific analysis, since a larger uncertainty is given by the underlying Poisson statistics ($\sigma_{Nph}=\sqrt{N_{ph}}\ge 1$); nonetheless, monitoring the position of the 1-photon peak would allow to keep under control the gain of the entire electronic chain and is therefore a desirable feature.

To decrease the required silicon surface, we are also investigating a new light guide design, to be added to the current Winston cone design or to replace it. One additional issue that needs to be addressed is the SiPM sensitivity to low-frequency photons: for IACT applications, sensitivity beyond the range of the Cherenkov spectrum largely increases the background due to the abundance of atmospheric photons. Finally, additional constraints on the pixel weight, power consumption, etc. must be considered to provide a valid PMT alternative.

\section{SiPM analog sum}

SiPM sensors are arrays of silicon diodes, operated under reverse bias in Geiger-counter mode: for each detected photon an element in the array fires, causing a current signal determined by the SiPM gain ($\sim 10^6$ electrons per photon), which is proportional to the operational voltage above the diode breaking voltage. The overall current signal is thus proportional to the number of firing cells, and therefore to the number of detected photons. The intrinsic linearity of a SiPM is one possible cause of concern: if too many photons impinge on the sensor at the same time the number of activated cells saturates. This is not a concern in our case, since the LST pixel must be linear up to a maximum of $\sim 2000$ photons impinging on the active area, which is far less than the expected number of SiPM cells, under any reasonable assumption on the cell size. 

In principle, the output of several SiPMs could be summed in a trivial way by connecting all sensors in parallel: each sensor is indeed a sum of individual cells, arranged in such a simple configuration. Except for considerations on the SiPM process yield, this approach would be equivalent to the manufacture of a single, large sensor. In practice, this would imply a huge detector capacitance proportional to the sensor's area, typically around $\sim 50$~pF$/$mm$^2$: at the input of an amplifier circuit, this would translate into an unacceptable level of electronic noise. Therefore, some method of summing the individual sensors while keeping a low input noise must be investigated. 

Currently, three avenues are being pursued to implement the analog sum of several SiPMs: a discrete component approach, based on a common-base amplifier design, a discrete component approach, based on a two-stage operational amplifier design, and a fully integrated approach, with a dedicated ASIC.

\begin{figure}
  \centering
  \includegraphics[width=0.9\textwidth]{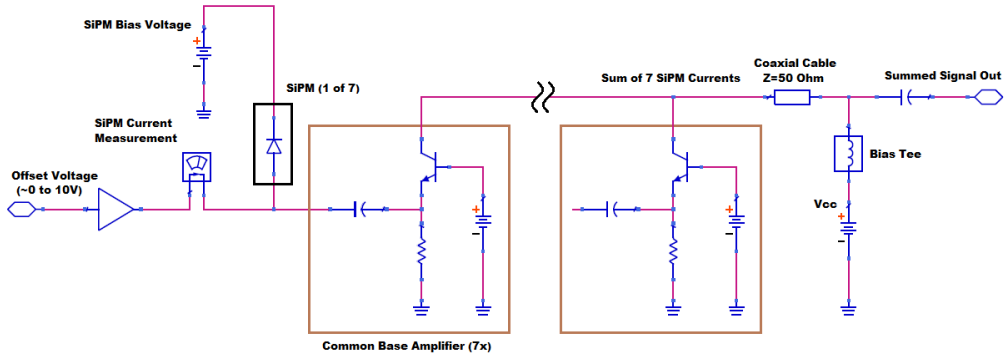}
  \caption{Adder circuit with common-base amplifiers. On the left, one SiPM sensor is shown, with the corresponding amplifier inside the brown box. In the current prototype this scheme is repeated seven times, and the total current drawn from the power line is read throught the capacitor on the right.}
  \label{fig:cmbase}
\end{figure}

A first sum scheme, developed at the Max Planck Institute for Physics, Munich, is shown in Fig.~\ref{fig:cmbase}. The signal from each individual sensor is amplified by means of a common-base transistor amplifier, to produce a current pulse of adequade magnitude. The total current pulse from all the amplifiers gives a voltage signal on the output capacitor proportional to the number of detected photons. Small differences in the gain of each sensor can be compensated by a small offset voltage applied on top of the baseline HV power supply. In the current prototype, 7 sensors are arranged in 3 sets (2, 3, and 2 sensors each), for which the operational voltage can be fine-tuned with a maximum swing of $\sim 10$~V. The gain adjustment system can also be used to power off malfunctioning sensors, or to partially turn off one pixel in case power consumption exceeds reasonable limits, e.g. in the case of operation in full-moon nights. In the current prototype the single-photon peak can be separated with a signal-to-noise ratio $\sim 4$; the power consumption is $\sim 300$~mW, SiPM power excluded, while the pulse width is $\sim 4$~ns.

\begin{figure}
  \centering
    \includegraphics[width=0.9\textwidth]{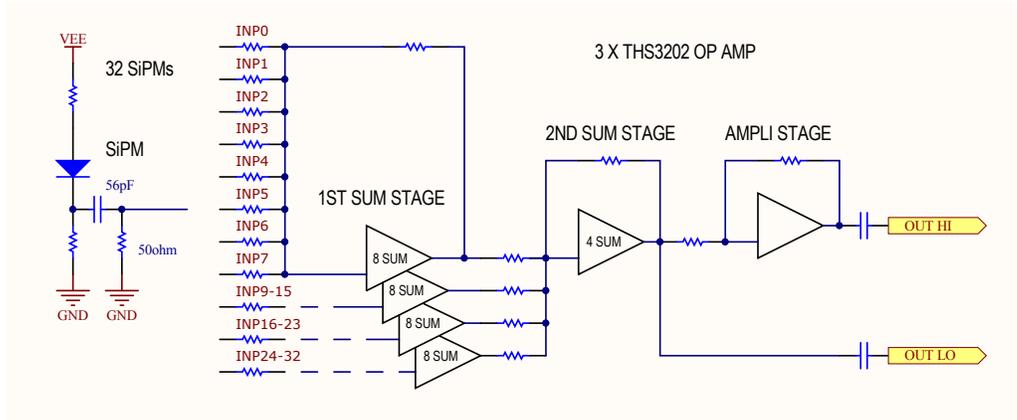}
    \caption{Adder circuit based on a two stage op-amp design. In the current prototype, the signals from eight SiPMs are filtered (high-pass) and summed into the first sum stage; four such first-stage sums are then summed into one final output in the second-stage adder. Finally, a further amplification stage is possible to split the output to have two gain levels.}
  \label{fig:opamp}
\end{figure}

A second prototype circuit for the analog sum, developed at the INFN laboratories in Padova, is shown in Fig.~\ref{fig:opamp}. A first sum stage based on a fast, low-noise operational amplifier collects and sums the signals from the SiPM sensors, with some gain. A second stage collects and sums signals from the first sum signals, to give a single output; some additional gain is applied here. Finally, one last additional gain stage can be used to implement a two-gain output design, e.g. like the one in the current LST camera design \cite{ref:pacta}. A 16 $3\times 3$-mm$^2$ SiPM prototype was built, demonstrating the capability of clearly separating the 1-photon peak with signal-to-noise ratio $\sim 3$; a 32 SiPM version is in production, together with a version mounting 9 $6\times 6$-mm$^2$ sensors. The power consumption for the 32-channel version is $\sim 300$~mW, SiPM power excluded, while the pulse width is $\sim 3$~ns. 

\begin{figure}
  \centering
  \begin{minipage}{0.4\textwidth}
    \centering
    \includegraphics[width=\textwidth]{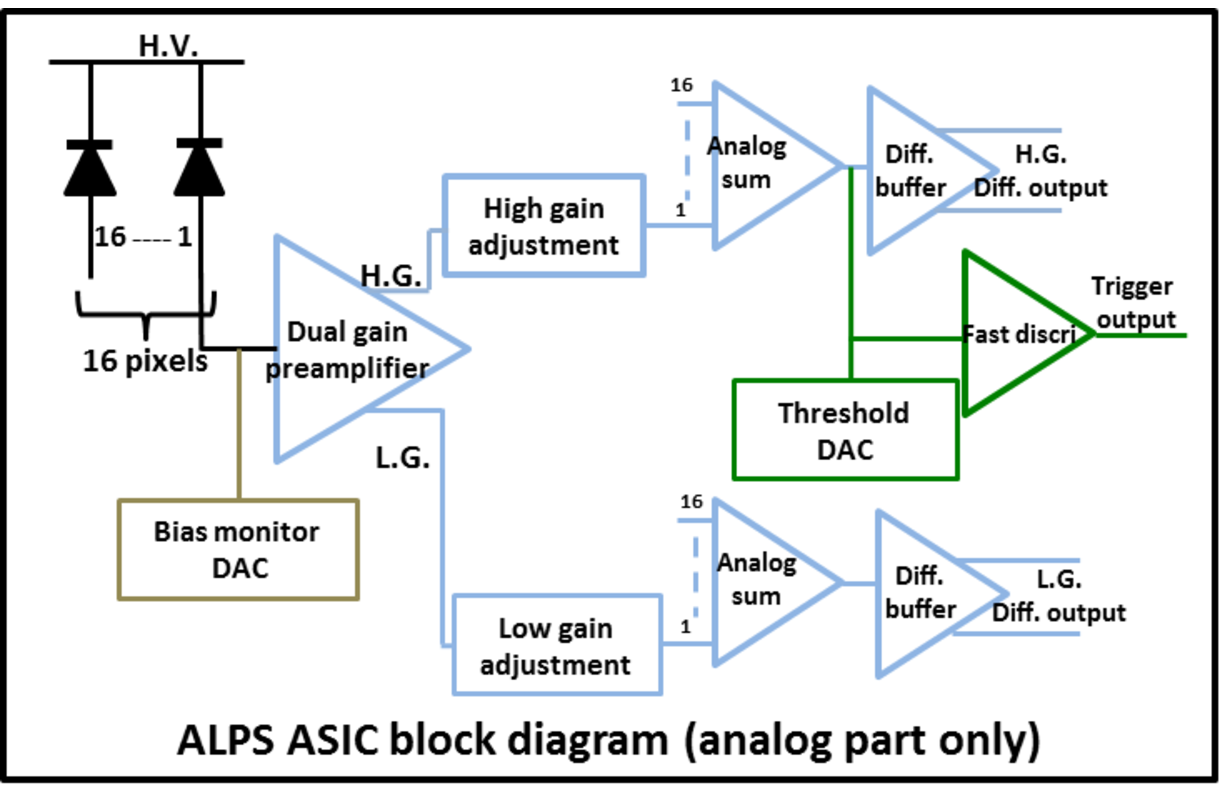}
  \end{minipage}%
\begin{minipage}{0.55\textwidth}
  \includegraphics[width=\textwidth]{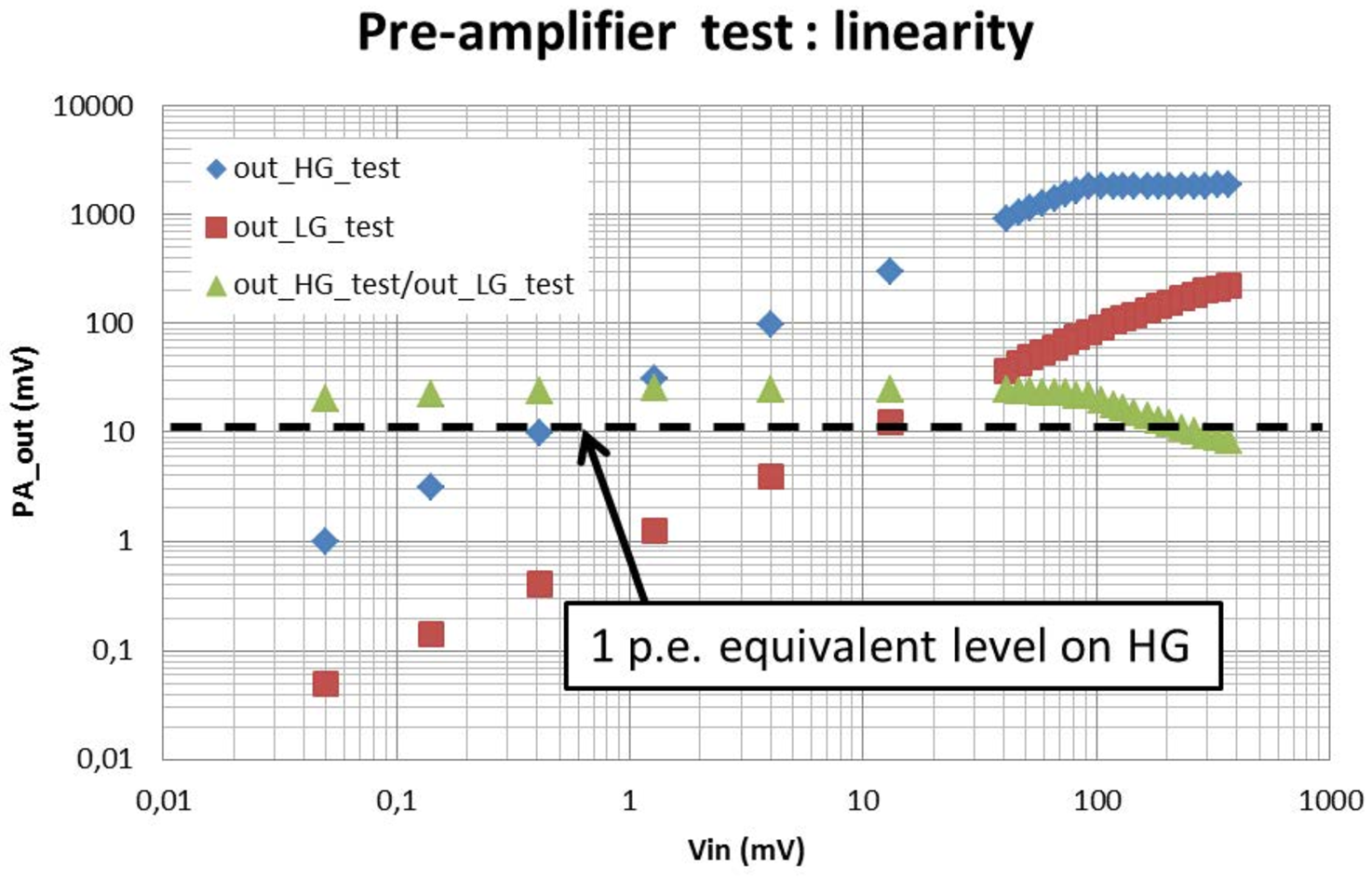}
\end{minipage}%
\caption{Left: ALPS block scheme. 16 input channels are individually amplified, adjusted for gain and summed, in both the low- and high-gain paths. Right: measurement of the gain and linearity of the preamplifiers.}
\label{fig:alps}
\end{figure}

A fully integrated approach is also pursued: an ASIC, labeled ``ALPS'', was developed at the IN2P3 laboratories at LAPP, Annecy. In Fig.~\ref{fig:alps}, left, the block scheme of ALPS is shown. Each of the 16 input channels is preamplified in a dual-gain preamplifier; the gain of each channel can be adjusted to compensate for a difference in gain among the SiPM sensors by means of digitally controlled resistors. The amplified signals are summed into the two outputs. Additional capabilities include a fast trigger output, based on a comparator with adjustable threshold. In Fig.~\ref{fig:alps}, right, the gain and linearity of the low- and high-gain outputs are shown, together with the output noise level. The power consumption is $<30$~mW, while the pulse width is $<5$~ns FWHM, and signal-to-noise ratio at the 1-phe level is $>5$. The first ASICs were received and electrically tested, confirming the simulated performance. Tests with a 16-sensor SiPM array are planned in the next future.

\section{Light guides}
The non-imaging reflecting-foil concentrators (``Winston cones'') designed for the PMT pixel are not optimal for SiPM sensors. The major issue is the incidence angle of photons on the sensor surface: a PMT has a curved entrance window, so a large incidence angle on the periphery is acceptable (actually, this increases quantum efficiency by some amount, due to possible double crossing of the photocathode). Large incidence angles on the flat layer of protective resin on top of the silicon plane in a SiPM sensor would lead instead to total reflection. 

\begin{figure}
  \centering
  \begin{minipage}{0.35\textwidth}
    \centering
    \includegraphics[width=\textwidth]{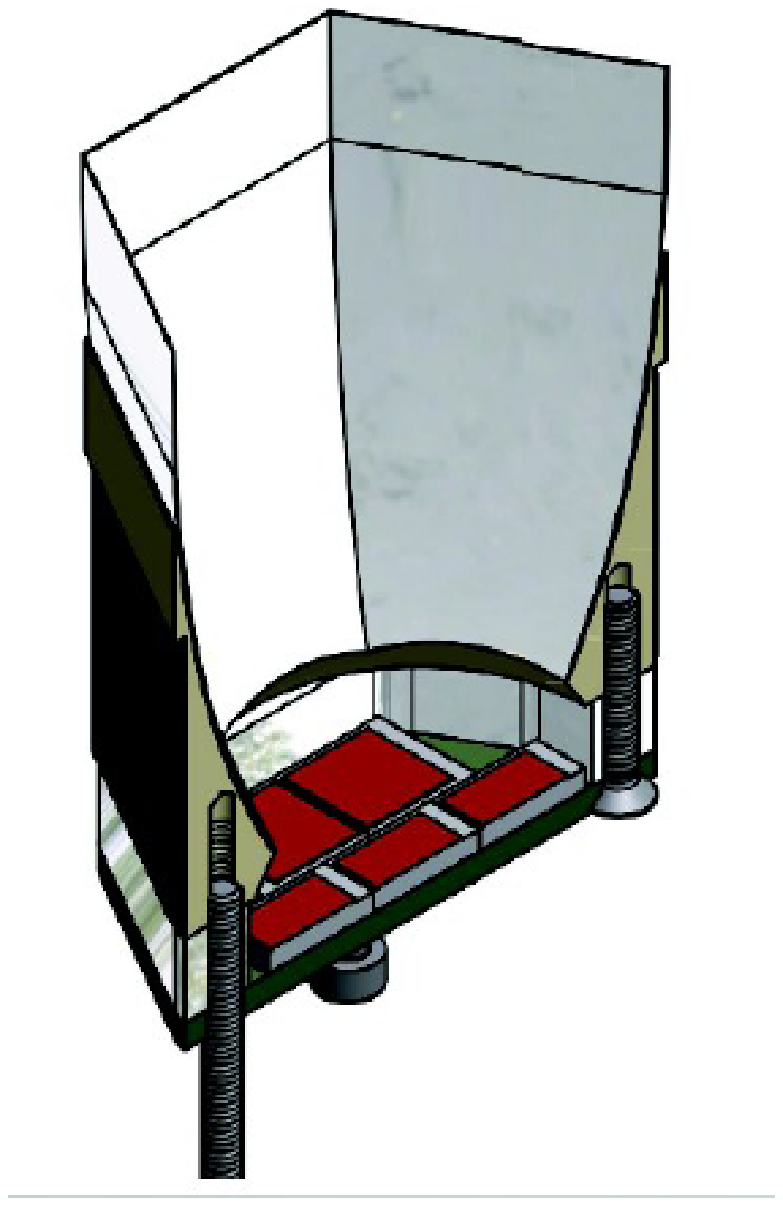}
  \end{minipage}%
\begin{minipage}{0.45\textwidth}
  \includegraphics[width=\textwidth]{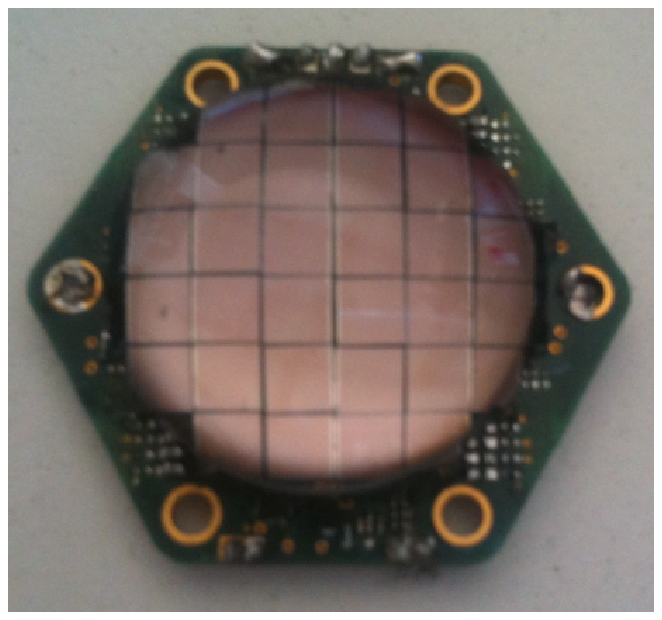}
\end{minipage}%
\caption{Left: SiPM pixel (MAGIC dimensions) with $6\times 6$~mm$^2$ sensors and a modified Winston cone, from which part of the back has been removed. Right: SiPM pixel (MAGIC dimensions) with with $3\times 3$~mm$^2$ sensors and a plane-convex lens on top.}
\label{fig:optics}
\end{figure}

This issue can be mitigated without redesigning the light guide from scratch. A first solution is to remove a small section from the back of the Winston cone with a transverse cut, since this is the part that causes the largest incidence angles on the sensor plane. Simulation of the optical system can be used to determine the thickness to remove, and the distance of the sensor plane from the exit aperture of the Winston cone, to ensure the maximum light collection (see Fig.~\ref{fig:optics}, left).
 A second solution consists of the addition of a plane-convex lens on top of the silicon sensors, to create a curved surface similar to the entrance window of PMTs (see Fig.~\ref{fig:optics}, right). Again, raytracing simulations are used to determine the profile of the lens that maximizes efficiency. An antireflecting coating can be applied to the lens, or even a red-filtering coating to reduce the detection of photons from the low-frequency, noise-dominated end of the visible spectrum.
Both approaches are being investigated, and prototypes are under test.

A complete redesign of the light guide is also being investigated. 
In particular, an effort to design a solid concentrator (plastic or glass) is ongoing: the additional concentration factor ensured by the high refraction index of the optical material would allow to decrease the amount of silicon needed in the sensor plane even further. In evaluating this solution, care will be taken to remain within the camera mass budget.

\section{The MAGIC cluster}

\begin{figure}
  \centering
    \includegraphics[width=0.9\textwidth]{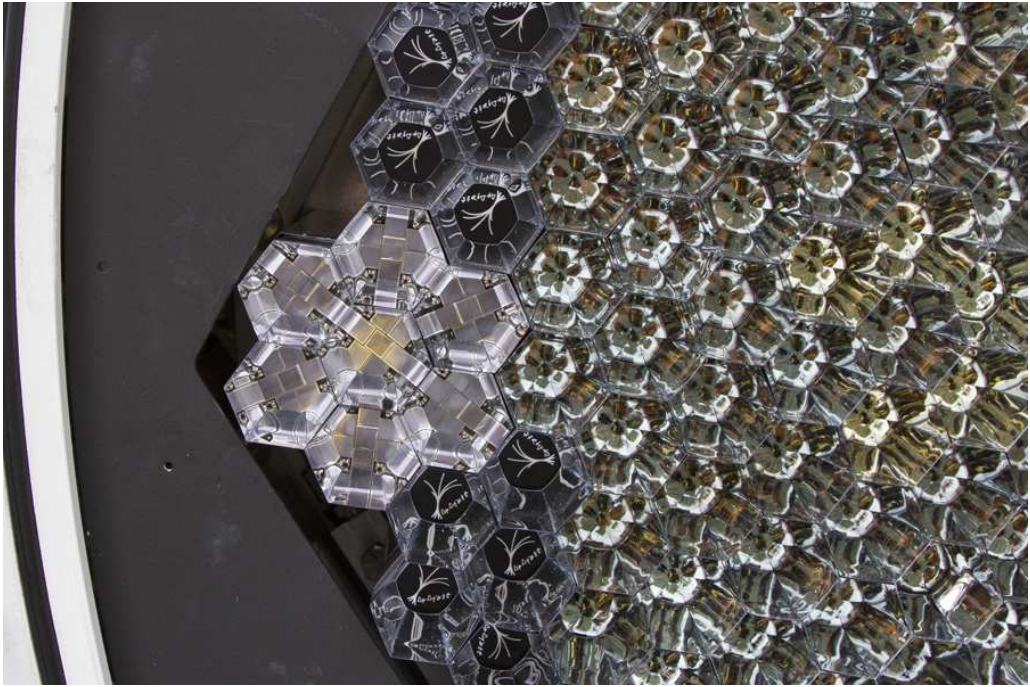}
    \caption{The installed MAGIC SiPM cluster (center) next to the PMT clusters (right half). One can also recognise some dummy pixels above and below the SiPM cluster.}
  \label{fig:magic_cluster}
\end{figure}

As an intermediate milestone, a smaller cluster designed to fit into the MAGIC camera structure \cite{ref:MAGICcam} is being developed. In this case the pixel area is smaller ($\sim5$~cm$^2$), the same size as the MAGIC PMTs), partially relieving the complexity in design, and the possibility of testing a prototype and validating each solution in a functional IACT is particularly exiting.
A first cluster is designed and assembled at the Max Planck Institute for Physics in Munich within the Otto Hahn research group. The design is based on the common-base adder scheme (for 7 SiPM sensors of $6\times6$~mm$^2$ size for each channel) and on a modified MAGIC light guide (see Fig.~\ref{fig:optics}, left). The adder is optimized to keep the shape of the individual SiPMs signals narrow (important for signal to noise ratio) and for a low power consumption. The design of the light guide is optimized to collect light from the entire mirror dish correcting for the incident angle acceptance of the silicon using non-sequential ray tracing software \cite{ref:robast}. Yet this first cluster is not optimized for high fill factors (dead space between individual sensors and fitting the output shape of the light guide) which will be a major task for the successor. The first cluster prototype passed intensive tests in Munich for stability and reliability of operation and then installed in La Palma in the MAGIC camera next to the PMT clusters in May 2015, see Fig.~\ref{fig:magic_cluster}. The obtained data is recorded by the regular MAGIC readout as the output signals of the SiPM cluster are identical to the ones from PMT cluster. Performance and long term tests are ongoing for a fair and detailed comparison between PMT and SiPM clusters,
which will be crucial for the decision on the next steps in SiPM cluster developments. 
To address the fill factor problem and to tests alternative SiPM sensors a second cluster is being assembled, based on the op-amp adder circuit (32 $3\times3$~mm$^2$ SiPM sensors) and on a curved lens behind the standard MAGIC light guide (see Fig. ~\ref{fig:optics}, right). At this time, assembly and installation are expected to complete within the end of 2015.

\section{Conclusion}
Several large IACT pixel prototypes are being built and tested, to demonstrate the feasibility of SiPM solutions for large IACT cameras. As an intermediate milestone, a smaller demonstrator unit has been completed and is being tested in realistic conditions in the camera of the MAGIC telescope; other alternative designs will be completed and tested within the year. R\&D will continue, aiming to a competitive pixel design for the CTA LST camera.

\section*{Acknowledgments}
We gratefully acknowledge support from the agencies and organizations listed under Funding Agencies at this website: http://cta-observatory.org/.

We acknowledge the cooperation of the MAGIC Collaboration, for giving us the possibility of testing the small-size pixel cluster in the MAGIC camera.

\end{document}